\documentclass[reprint,amsmath,amssymb,prl,aps]{revtex4-2}

\usepackage{graphicx}
\usepackage{dcolumn}
\usepackage{bm}
\usepackage{color}

\newcommand{\mnras}{Mon.~Not.~Roy.~Astron.~Soc.}
\newcommand{\jcap}{JCAP}
\newcommand{\pasj}{Publ.~Astron.~Soc.~Japan}

\newcommand{\aap}{Astronomy~\&~Astrophysics}

\newcommand{\araa}{ARA\&A}
\newcommand{\pasp}{PASP}
\newcommand{\aapr}{AAPR}

\newcommand{\planck}{{\it Planck}~}
\newcommand{\hiMsun}{h^{-1}{\rm M}_\odot}
\newcommand{\hiMpc}{h^{-1}{\rm Mpc}}

\begin{document}

\preprint{APS/123-QED}

\title{First Identification of a CMB Lensing Signal Produced by $1.5$ Million Galaxies at $z\sim$4:\\
Constraints on Matter Density Fluctuations at High Redshift}

\author{Hironao Miyatake$^{1,2,3,4}$,Yuichi Harikane$^{5,6}$, Masami Ouchi$^{7,5,4}$, Yoshiaki Ono$^{5}$, Nanaka Yamamoto$^{3}$,\\ Atsushi J. Nishizawa$^{8,2,3}$, Neta Bahcall$^{9}$, Satoshi Miyazaki$^{7}$, Andr\'{e}s A. Plazas Malag\'{o}n$^{9}$}
\affiliation{$^1$Kobayashi-Maskawa Institute for the Origin of Particles and the Universe (KMI),
Nagoya University, Nagoya, 464-8602, Japan} 
\affiliation{$^2$Institute for Advanced Research, Nagoya University, Nagoya, 464-8601, Japan}
\affiliation{$^3$Division of Physics and Astrophysical Science, Graduate School of Science, Nagoya University, Furo-cho, Chikusa, Nagoya 464-8602, Japan}
\affiliation{$^4$Kavli Institute for the Physics and Mathematics of the Universe (Kavli IPMU, WPI), UTIAS, The University of Tokyo, 5-1-5 Kashiwa-no-ha, Kashiwa, Chiba 277- 8583, Japan}
\affiliation{$^5$Institute for Cosmic Ray Research, The University of Tokyo, 5-1-5 Kashiwanoha, Kashiwa, Chiba 277-8582, Japan}
\affiliation{$^6$Department of Physics and Astronomy, University College London, Gower Street, London WC1E 6BT, UK}
\affiliation{$^7$National Astronomical Observatory of Japan, 2-21-1 Osawa, Mitaka, Tokyo 181-8588, Japan}
\affiliation{$^8$DX Center, Gifu Shotoku Gakuen University, Takakuwa-Nishi, Yanaizucho, Gifu, 501-6194, Japan}
\affiliation{$^9$Department of Astrophysical Sciences, Peyton Hall, Princeton University, Princeton, NJ 08540, USA}

\date{\today}

\begin{abstract}
We report the first detection of the dark matter distribution around Lyman break galaxies (LBGs) at high redshift through the Cosmic Microwave Background (CMB) lensing measurements with the public \planck PR3 $\kappa$ map. The LBG sample consists of 1,473,106 objects with the median redshift of $z \sim 4$ that are identified in a total area of 305 deg$^2$ observed by the Hyper Suprime-Cam (HSC) Strategic Survey Program (SSP) survey. After careful investigations of systematic uncertainties, such as contamination from foreground galaxies and Cosmic Infrared Background (CIB), we obtain the significant detection of the CMB lensing signal at $5.1\sigma$ that is dominated by 2-halo term signals of the LBGs. Fitting a simple model consisting of the Navarro–Frenk–White (NFW) profile and the linear-bias model, we obtain the typical halo mass of $M_{\rm h}=2.9^{+9.5}_{-2.5}\times10^{11}\ \hiMsun$
. Combining the CMB lensing and galaxy-galaxy clustering signals on the large scales, we demonstrate the first cosmological analysis at $z\sim4$ that constrains $(\Omega_{{\rm m}0}$, $\sigma_8)$. We find that our constraint on $\sigma_8$ is roughly consistent with the \planck cosmology, while this $\sigma_8$ constraint is lower than the \planck cosmology over the $1\sigma$ level. This study opens up a new window for constraining cosmological parameters at high redshift by the combination of CMB and high-$z$ galaxies as well as studying the interplay between galaxy evolution and larges-scale structure at such high redshift, by upcoming CMB and optical and near-infrared imaging surveys. 
\end{abstract}

\maketitle

\section{Introduction}
\label{sec:intro}
Understanding the interplay between galaxy evolution and large-scale structure is key to understanding cosmic evolution. Since galaxies are formed in dark matter halos through gas cooling, such an interplay can be studied by measuring the connection between dark matter halos and galaxies \cite[see reviews by][]{Somerville:2015,Wechsler:2018}. Galaxy-galaxy clustering, the auto-correlation of galaxy positions, is one of the statistical probes that has been widely used to investigate the galaxy-halo connection \cite[e.g.,][]{McCracken:2015,Hatfield:2016,Ishikawa:2017,Cowley:2018,Harikane2018}. Galaxy-galaxy lensing, the cross-correlation between galaxy positions and weak lensing shear of background galaxies, is rapidly emerging as another powerful probe because it enables the direct measurement of dark matter distribution around galaxies \cite[e.g.,][]{Leauthaud:2012,Coupon:2015,Huang:2020,Taylor:2020}.

The combined measurement of galaxy-galaxy clustering and galaxy-galaxy lensing allows us to constrain cosmological parameters through growth of structure, and thus it is now the standard technique in cosmological galaxy surveys. Such studies can be divided into those using both small and large scales which enable us to investigate the galaxy-halo connection simultaneously \cite[e.g.,][]{Cacciato:2013,More:2015}, and those focusing on large scales to avoid possible systematics due to our incomplete understanding of galaxy physics \cite[e.g.,][]{Mandelbaum:2013,More:2015,DES:2018,Heymans:2021}. These measurements advanced our understanding of cosmic evolution of large-scale structure at $z<1$ through the constraint on the amplitude of matter power spectrum $\sigma_8$, which are consistent with the \planck cosmology so far. As an independent test, it is of great importance to add such a measurement at higher redshift. Having structure growth measurements with a wide redshift may also allow us to address the $H_0$ tension between measurements of late time probes such as the type Ia supernova distance and early time probes such as cosmic microwave background (CMB) \cite[e.g.,][]{Riess:2019, DiValentino:2021}, through the sensitivity to early dark energy\cite{Klypin:2021}.

Recently, detailed study results of galaxy properties and galaxy-halo connection through clustering measurement of Lyman break galaxies (LBGs) at $z\sim4-7$ are reported \cite{Hildebrandt:2009,Harikane2018}, where they used wide-field galaxy imaging survey data such as the Canada- France-Hawaii Telescope Legacy Survey (CFHTLS) and the Subaru Hyper Suprime-Cam (HSC) Subaru Strategic Program (SSP) Survey \cite{Miyazaki:2018}. Although one of the next steps is measuring weak lensing on background galaxies around the LBGs, it is practically impossible due to the lack of data for shear measurements of galaxies behind LBGs. Instead, we can use weak lensing on the CMB anisotropies, which is sensitive to the large-scale structure at $z>1$. In this paper, using CMB lensing measurements in the \planck satellite data, we present the first detection of dark-matter distribution around LBGs at $z\sim4$. We also show cosmological constraints from clustering and lensing measurements.

\section{Data}
\label{sec:data}
\subsection{HSC GOLDRUSH Catalog}
\label{sec:hsc}
We use the second public data release (PDR2) product taken by the HSC SSP from 2014 March to 2018 January \cite{Aihara2019}. We use the Wide layer data whose total full color full depth area is 305 $\mathrm{deg^2}$. The HSC data are reduced by the HSC SSP collaboration with hscPipe version 6.7 \cite{Bosch2018,Bosch:2019}. The typical $5\sigma$ limiting magnitudes in a $1.5''$ diameter aperture are 26.4, 25.9, 25.8, 25.1, and 24.2 mag in $g$, $r$, $i$, $z$, and $y$-bands, respectively.

To construct a $z\sim4$ LBG catalog of Great Optically Luminous Dropout Research Using Subaru HSC (GOLDRUSH) \cite{Ono2018}, we use the dropout selection method. Here we briefly describe the selection method; full details of the catalog construction is presented in \citet{Harikane:2021}.
We use the following color criteria to select $z\sim4$ galaxies:
\begin{eqnarray}
g-r &>& 1.0\\
r-i &<& 1.0\\
g-r &>& 1.5(r-i) + 0.8,
\end{eqnarray}
which are the same as the GOLDRUSH catalog construction with the first HSC data release \cite{Ono2018, Harikane2018, Toshikawa2018}. In addition to these criteria, we apply a detection limit of a $>5\sigma$ level in the $i$-band. We evaluate source detections with aperture magnitudes, and measure total fluxes and colors of sources with {\tt convolvedflux} magnitudes, which show good agreement with the {\tt cModel} magnitude in the first public data release \cite{Aihara2018}. We mask out regions affected by bright object halos or unreliable background subtractions. Finally, we select a total of $\sim$2,000,000 LBGs. \citet{Ono2018} used spectroscopically identified high-$z$ galaxies in the dropout sample to estimate the average redshift of the sample $\left < z \right > =3.8$ (median redshift of $z\sim 4$), which we adopt for our analysis. The surface number density and angular correlation functions of the LBGs agree well with previous studies. In our analysis, we use $1,643,863$ LBGs whose total $i$-band magnitudes are brighter than 25.5 mag, which is reduced to 1,473,106 LBGs after applying the \planck mask described below. Some foreground objects such as red galaxies at intermediate redshift can satisfy our selection criteria by photometric errors. A contamination rate is estimated to be $f_{\rm cont}\sim0.25$ based on HSC WIDE layer results obtained by \cite{Ono2018} who derive a fraction of contaminating galaxies as a function of $i$-band magnitude by calculating the number of foreground galaxies that satisfy the color criteria.

\subsection{\planck $\kappa$ map}
\label{sec:planck}
We construct the convergence map using lensing products in the third public data release from the \planck  satellite mission (PR3) \cite{PlanckLensing2018} \footnote{https://pla.esac.esa.int}. We use the minimum variance (MV) estimator and masks in the baseline lensing potential estimates. The MV estimators are provided as harmonic coefficients with the maximum multipole $L_{\rm max}=4096$, while the mask is in real space. To construct the convergence map, we subtract the MV mean field from the MV convergence, and transform the harmonic coefficients to a real-space map applying a Gaussian filter with $\sigma=2^{\prime}$. We apply the mask to remove the galactic plane, point sources, and Sunyaev-Zel'dovich (SZ) clusters, which leaves a total unmasked sky fraction $f_{\rm sky}=0.671$. This results in the healpix map with \texttt{NSIDE=2048} where the area of each healpixel is 2.95~$\mathrm{arcsec^2}$. Note that the HSC-SSP survey does not include the galactic plane, and that we have a full overlap between our LBG sample and the \planck convergence map. After removing regions of point sources and SZ clusters masked in the \planck data, the total overlap area is 270~$\mathrm{deg^2}$.

\section{Measurement}
\label{sec:measurement}
To measure the stacked convergence profile, we first create galaxy-centric angular bins for each galaxy by linearly dividing $0^\prime<\theta<20^\prime$ into 14 bins, and identify healpixels that belong to each bin. Note that we validate the binning scheme as follows. First, we generate the 10000 realizations of mock signals by fluctuating the convergence signal following the jackknife covariance described below. We then compute $\chi^2=\sum_{\rm ij}(\kappa_{\rm mock, i}-\kappa_i)\rm {Cov}^{-1}_{\rm ij}(\kappa_{\rm mock, j}-\kappa_j)$ for each mock realization and confirmed that the $\chi^2$ distribution follows the predicted $\chi^2$ distribution with degree of freedom of 14.

We then collect the healpixels over all galaxies, and compute the average convergence in each bin. To subtract the residual mean field, we also perform the same measurement around random points, and subtract the random signal from the observed convergence profile. The number of random points is 40 times larger than that of the LBGs. 

We estimate the jackknife covariance dividing the PDR2 area into 262 subregions such that each subregion has a similar number of galaxies, and measure the mean-subtracted signal for each jackknife subregion. Note that the covariance is highly correlated due to the low-pass filter applied in the process of MV convergence estimation.

To test the contamination from cosmic infrared background (CIB), we stack \planck SZ-nulled temperature map ({\tt SMICA\_noSZ}) around the LBGs. We do not find a gradient in the signal, and thus conclude that there is no significant CIB contamination. In addition we confirm that there is no significant B-mode signal.

\section{Results}
\label{fig:results}
\begin{figure}
\includegraphics[width=0.95\columnwidth]{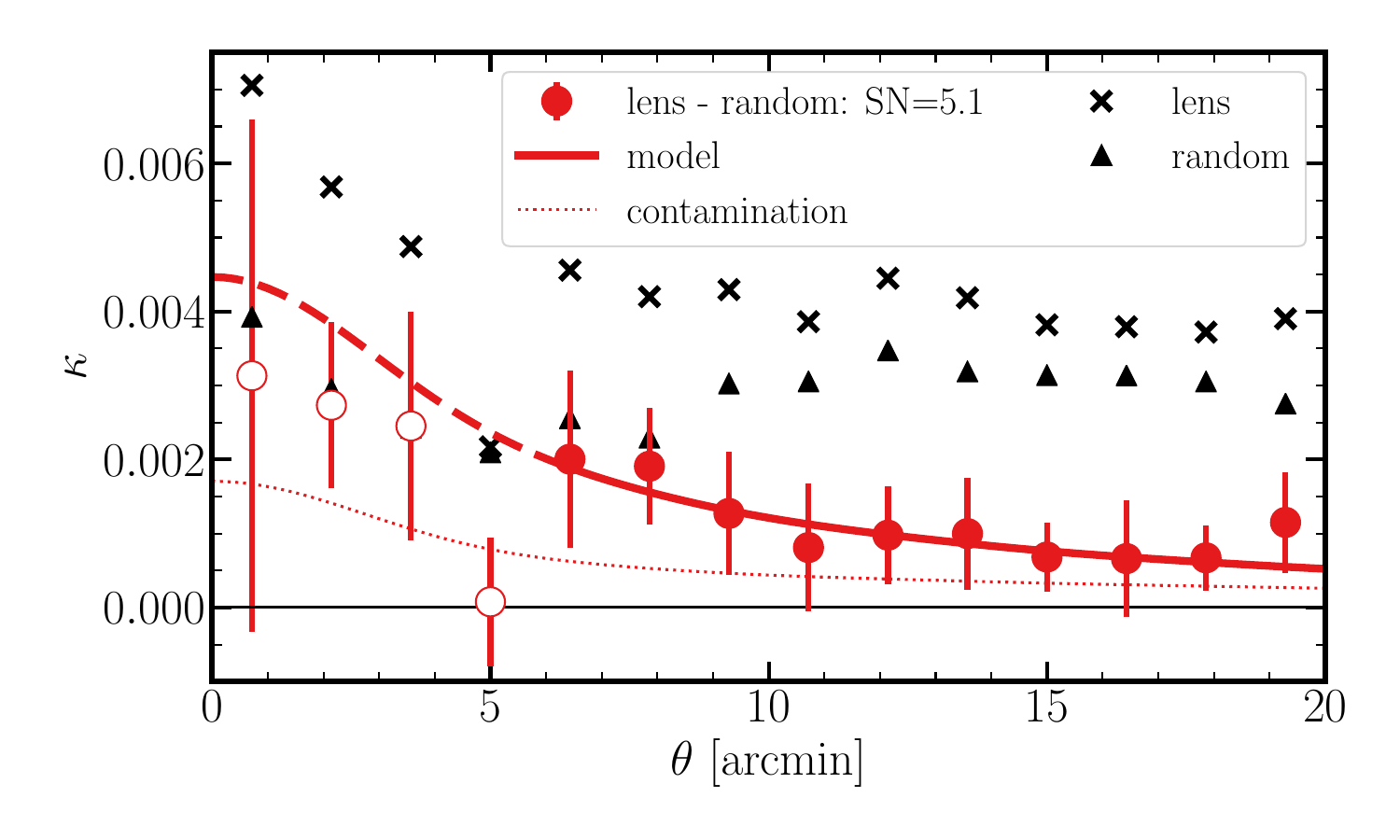}
\caption{\planck CMB lensing measurements around $z\sim4$ LBGs. The red circles denote the measured lensing signal after the mean-field subtraction, where the filled (open) circles denote data points used for (excluded from) our model fit. The black crosses (triangles) denote the lensing signal before the mean-field subtraction (mean-field signal measured by stacking the \planck $\kappa$ map around random points). The red solid (dashed) line represents our best-fit model of the lensing signal at the scales used for (excluded from) our model fit. The red dotted line shows the lensing signal of low-redshift contaminating galaxies in the LBG sample constrained by optical lensing measurements with the HSC shapes (for details, see text). The significance of the signal is 5.1$\sigma$ (3.5$\sigma$) against null (the contamination signal).}
\label{fig:profile}
\end{figure}

In Fig.~\ref{fig:profile}, we show the measured convergence profile together with the mean-field profile and the convergence profile before the mean-field subtraction. We confirm that the observed amplitude of mean-field signal is typical by measuring the convergence around random points using the publicly-available simulated lens map in PR3. Defining the significance of the convergence profile after the mean-field subtraction as $[\rm{S/N}]_\kappa=\sum_{ij} \kappa_i {\rm Cov}_{ij}^{-1} \kappa_j$, we obtain the significance level of $[\rm{S/N}]_\kappa=5.1$ with all radial bins. Note that the convergence of each sub-field of the PDR2 data \cite{Aihara2019} tends to be positive but we do not find a clear signal. This is because these sub-fields have a much smaller number of LBGs compared to our full sample.

The convergence profile is contaminated by foreground galaxies as described in the previous section. We thus model our signal as $\kappa^{\rm obs}_{\rm tot} = (1-f_{\rm cont})\kappa^{\rm obs}_{\rm LBG}+f_{\rm cont}\kappa^{\rm obs}_{\rm cont}$. We use the halo models for the convergence profiles of real LBGs and contaminating galaxies with different halo masses and redshifts.

Let us first consider the convergence profile without the \planck beam. In the halo model approach, the convergence profile is composed of 1-halo and 2-halo terms; $\kappa(\theta) = \kappa_{\rm 1h}(\theta)+\kappa_{\rm 2h}(\theta)$, where $\kappa_{\rm 1h}(\theta)$ and $\kappa_{\rm 2h}(\theta)$ are the contribution from dark matter halos around galaxies and neighboring halos, respectively. We model the 1-halo term as $\kappa_{\rm 1h}(R) = \Sigma_{\rm cr}^{-1}(z_{\rm l}, z_{\rm CMB}) \bar{\rho}_{\rm m} \int (k{\rm d}k/2\pi)\tilde{u}_{\rm m}(k;M_{\rm h}, z_{\rm l})J_0(kR)$, where $z_{\rm l}$ ($z_{\rm CMB}$) is the redshift of lens (CMB), $\bar{\rho}_{\rm m}$ is the mean matter density at present, $\tilde{u}_m(k;z_{\rm l})$ is the Fourier-space Navarro–Frenk–White (NFW) \cite{Navarro:1996, Navarro:1997} profile truncated at the radius where the enclosed mass is 200 times the mean density $R_{\rm 200}$, $J_0(x)$ is the zeroth-order Bessel function of the first kind. We adopt the analytical form of $\tilde{u}_m(k;z_{\rm l})$ provided by \citet{Takada2003}, and assume the concentration-mass relation derived by \citet{Duffy2008}. The critical surface density is defined as $\Sigma_{\rm cr}(z_{\rm l}, z_{\rm s})=c^2 [4\pi G]^{-1} d_{\rm A}(z_{\rm s}) [d_{\rm A}(z_{\rm l})d_{\rm A}(z_{\rm l},z_{\rm s})]^{-1} (1+z_{\rm l})^{-2}$, where $c$ is the speed of light, $G$ is the gravitational constant, $d_{\rm A}(z_{\rm s})$, $d_{\rm A}(z_{\rm l})$, and $d_{\rm A}(z_{\rm l}, z_{\rm s})$ are the angular diameter distances for the lens-source system. To model the 2-halo term, we adopt the linear bias model, i.e., $\kappa_{\rm 2h}(R)=\Sigma_{\rm cr}^{-1}(z_{\rm l}, z_{\rm CMB}) b(M_{\rm h}, z_{\rm l}) \bar{\rho}_{\rm m} \int (k{\rm d}k/2\pi) P_{\rm m}(k, z_{\rm l})J_0(kR)$, where $b(M_{\rm h}, z_{\rm l})$ is the linear bias parameter, and $P_{\rm m}(k, z_{\rm l})$ is the linear matter power spectrum at $z_{\rm l}$. We adopt the fitting function for the linear bias derived by \citet{Tinker2010}, and compute the matter power spectrum using \texttt{CAMB} \cite{Lewis2002}. Note that in our model the 1-halo and 2-halo terms are connected through halo mass. Finally, we convolve the \planck beam by applying the low pass filter $L\leq4096$ and the Gaussian filter with $\sigma=2^{\prime}$ to obtain a model for the observed signal $\kappa^{\rm obs}(\theta)$.

To estimate the halo mass of contaminating galaxies, we measure the weak lensing signal around the LBGs with the HSC first-year shear catalog \cite{Mandelbaum2018a,Mandelbaum2018b}. This measurement enables us to pick up the lensing signal from contaminating galaxies, since most of the HSC source galaxies reside at $z<4$ and thus are insensitive to the real LBGs at $z=3.8$. We first measure the excess surface mass density $\Delta\Sigma(R)$, following the procedure described in \citet{Mandelbaum2018a}. Note that we include the random subtraction in the estimator, and we do not apply the boost factor correction since the boost factor is less than $1\%$ \cite{Mandelbaum:2005}. Since we use the full probability density functions of photometric redshift (photo-$z$), the photo-$z$ correction needed for the use of point estimates is not applied \cite{Nakajima:2012}. The covariance is estimated using the jackknife method using the same jackknife subsamples as the CMB lensing measurement. We fit the signal with a model based on halo occupation distribution (HOD) with off-centering (5 parameters for HOD and 2 parameters for off-centering), which is similar to the one described in \citet{More:2015}, using the publicly-available Markov Chain Mote Carlo (MCMC) sampler called \texttt{emcee} \cite{Foreman-Mackey2013}. We translate the best-fit model in $\Delta\Sigma$ into convergence $\kappa$ to obtain the contamination signal \cite[see Eq.(85) in][]{Umetsu:2020}. We estimate the uncertainty in the contamination signal by generating the convergence profile using the MCMC chains. The 1-sigma uncertainty is at most $25$\% of the contamination signal itself, which is much smaller than the statistical uncertainty of our CMB lensing measurements, as can be seen in Fig~\ref{fig:profile}. We thus conclude that the uncertainty of contamination signal does not significantly impact our analysis.

We then fit the convergence model to the signal varying the halo mass of LBGs.We use the signals at $6^{\prime}<\theta<20^{\prime}$ to remove small scales that can possibly suffer from residual signals at $L>2048$ \cite{PlanckLensing2018}. We show the best-fit model in Fig.~\ref{fig:profile} together with the expected contamination signal based on the halo mass derived above. Note that the contamination level is much lower than the observed signal because of the insensitivity of CMB lensing at low redshift. In fact, the significance of the observed signal against the contamination signal is $3.5\sigma$. The LBG halo mass is constrained as $M_{\rm h}=2.9^{+9.5}_{-2.5}\times10^{11}\ \hiMsun$ (68\% credible level) \footnote{Throughout this paper, we adopt the mode of the posterior distribution to infer the central value of the parameter, and the highest density interval of the posterior to infer the credible interval of the parameter. We use the marginalized posterior distribution when there are multiple fitting parameters}, which is consistent with previous studies \cite{Hildebrandt:2009,Harikane2018} where they derived the halo mass through clustering measurements. Note that the signal is dominated by the 2-halo term, as expected from the fact that $R_{\rm 200}$ corresponding to this halo mass is at $\theta\sim 0.02^\prime$ and the 1-halo term is almost completely smeared out by the Planck beam, and thus the halo mass constraint comes from the mass-bias relation by \citet{Tinker2010}.

\section{Discussions}
\label{sec:discussion}
A measurement of the matter correlation function can constrain cosmological parameters such as the matter energy density at present $\Omega_{{\rm m}0}$ and the amplitude of the matter power spectrum $\sigma_8$. Since galaxies are a biased tracer of the underlying matter distributions, we cannot directly measure the matter correlation function from the spatial distributions of galaxies (galaxy clustering). The matter correlation function, however, can be recovered by combining the galaxy clustering and galaxy lensing signal \cite{Baldauf2010}, because the clustering and lensing signal depend on the bias in a different manner. In the limit of large scales where the galaxy density fluctuation follows the linear bias model $\delta_{\rm g}=b\delta_{\rm m}$, the clustering signal provides the galaxy-galaxy correlation function $\xi_{\rm gg}=b^2\xi_{\rm mm}$, while the lensing signal indicates the galaxy-matter correlation function $\xi_{\rm gm}=b\xi_{\rm mm}$.

Adding a projected clustering measurement of the LBG sample \cite{Harikane:2021} to our lensing signal measurement, we demonstrate the first cosmological analysis through the growth of large-scale structure at $z\sim4$. We measure the projected clustering signal and its jackknife covariance of the LBGs in the same manner as the previous measurements \cite{Harikane2018}. To ensure large-scale signals are used for cosmological constraints, we use the clustering signal at $1.5^\prime < \theta < 6.5^\prime$ and lensing signal at $6.0^\prime < \theta < 20^\prime$, which corresponds to the comoving scale ranges of $2.4 < R/[\hiMpc] <9.4$ and $8.7 < R/[\hiMpc] < 30$, respectively, at $z=3.8$.

We estimate the impact of the uncertainty associated with LBG redshifts by using a subsample of LBGs which have spectroscopic redshifts. Using the spectroscopic redshifts, we find the model varies by $4\%$ and $17\%$ for lensing and clustering, respectively. These deviations are much smaller than the statistical uncertainties of the measurements which are typically $70\%$ and $40\%$ for lensing and clustering, respectively.

We constrain the cosmological parameters $\Omega_{{\rm m}0}$ and $\sigma_8$ by simultaneously fitting the large scale clustering and lensing signal, while the other cosmological parameters are fixed to the \planck cosmology \cite{PlanckCosmology2018}. We model the lensing signal in the same manner as the modeling procedures described in the previous section with the following modifications. For the convergence profile of LBGs, we use only the 2-halo term and vary $b$ as a fitting parameter. We also vary $\Omega_{{\rm m}0}$ and $A_s$ on which the linear power spectrum depends. Both 1-halo and 2-halo term in the lensing signal from contaminating galaxies are fixed. We model the observed clustering signal as $\omega_{\rm tot}(\theta)=(1-f_{\rm cont})^2\omega_{\rm LBG}(\theta)$, where $\omega_{\rm LBG}(\theta)$ is the clustering signal from LBGs. Note that we ignore the contamination signal for clustering signal because it is suppressed by $f_{\rm cont}^2\sim0.06$. The projected clustering signal pf LBGs is modeled as $\omega_{\rm LBG}(\theta)=\int dz N^2(z)(dr/dz)^{-1}\int dk (k/2\pi)b^2P_{\rm mm}(k,z;\Omega_{{\rm m}0}, A_s)\\J_0(r(z)\theta k)$, where $N(z)$ is the normalized redshift distribution of galaxies taken from \cite{Ono2018}.

\begin{figure}
\includegraphics[width=\columnwidth]{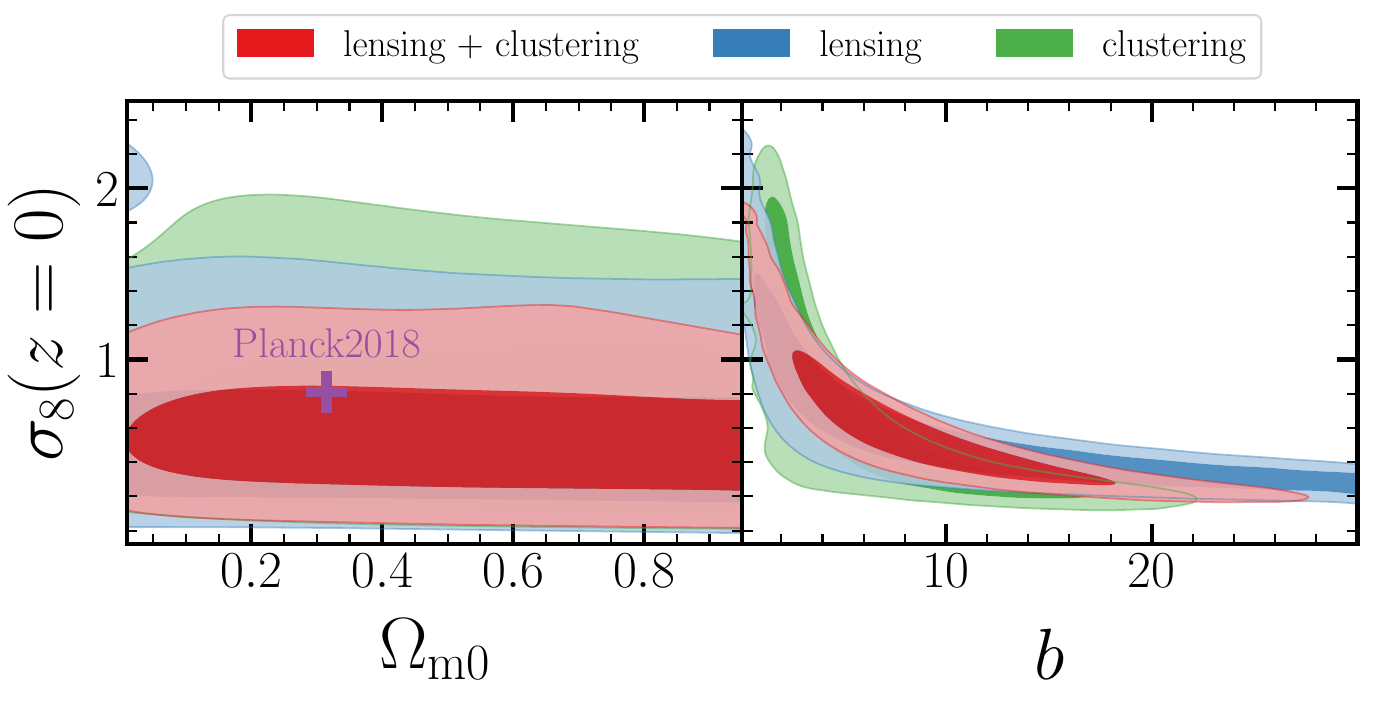}
\caption{Cosmological constraints from CMB lensing and galaxy-galaxy clustering measurements at $z\sim4$ (red contours). The blue (green) contours show constraints from the lensing (clustering) probe alone. The degeneracy between $\sigma_8$ and $b$ is reduced by the combination of these probes. These constraints are roughly consistent with the \planck cosmology \cite[purple cross;][]{PlanckCosmology2018}.}
\label{fig:cosmological_constraint}
\end{figure}

We perform an MCMC analysis with flat priors, $\Omega_{\rm m,0} \in [0.01, 0.95]$, $\ln 10^{10}A_s \in [0.1, 5.0]$, and $b \in [0.1, 30]$. Fig.~\ref{fig:cosmological_constraint} shows constraints on these parameters for the combined lensing and clustering results, together with constraints based on the lensing or clustering results alone. Since $\sigma_8$ and $b$ degenerate in a different manner for lensing and clustering at small $b$, the combination of lensing and clustering probes improves the constraint on $\sigma_8$, yielding $\sigma_8 = 0.46^{+0.25}_{-0.20}$ and $b = 6.5^{+5.2}_{-4.2}$ (68\% credible level). Although it is known that there is a strong degeneracy between $\Omega_{{\rm m}0}$ and $\sigma_8$ in a combined analysis of lensing and clustering \cite[e.g.,][]{More:2015, DES:2018, Heymans:2021}, the degeneracy is not observed
in our analysis, and there is almost no constraint on $\Omega_{\rm m0}$. This is due to the fact that the sensitivity of matter power spectrum to $\Omega_{\rm m0}$ significantly decreases at high redshifts.

\begin{figure}
\includegraphics[width=\columnwidth]{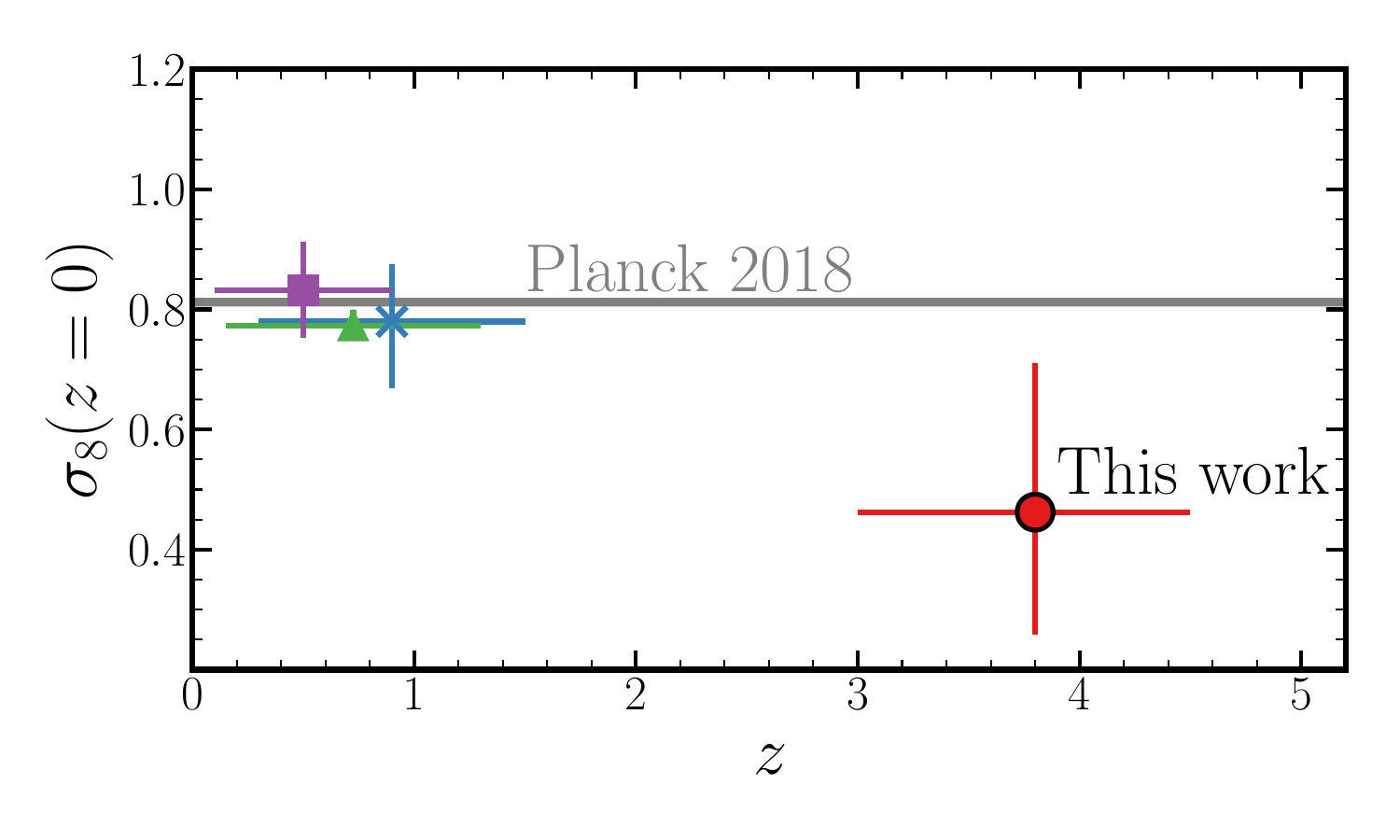}
\includegraphics[width=\columnwidth]{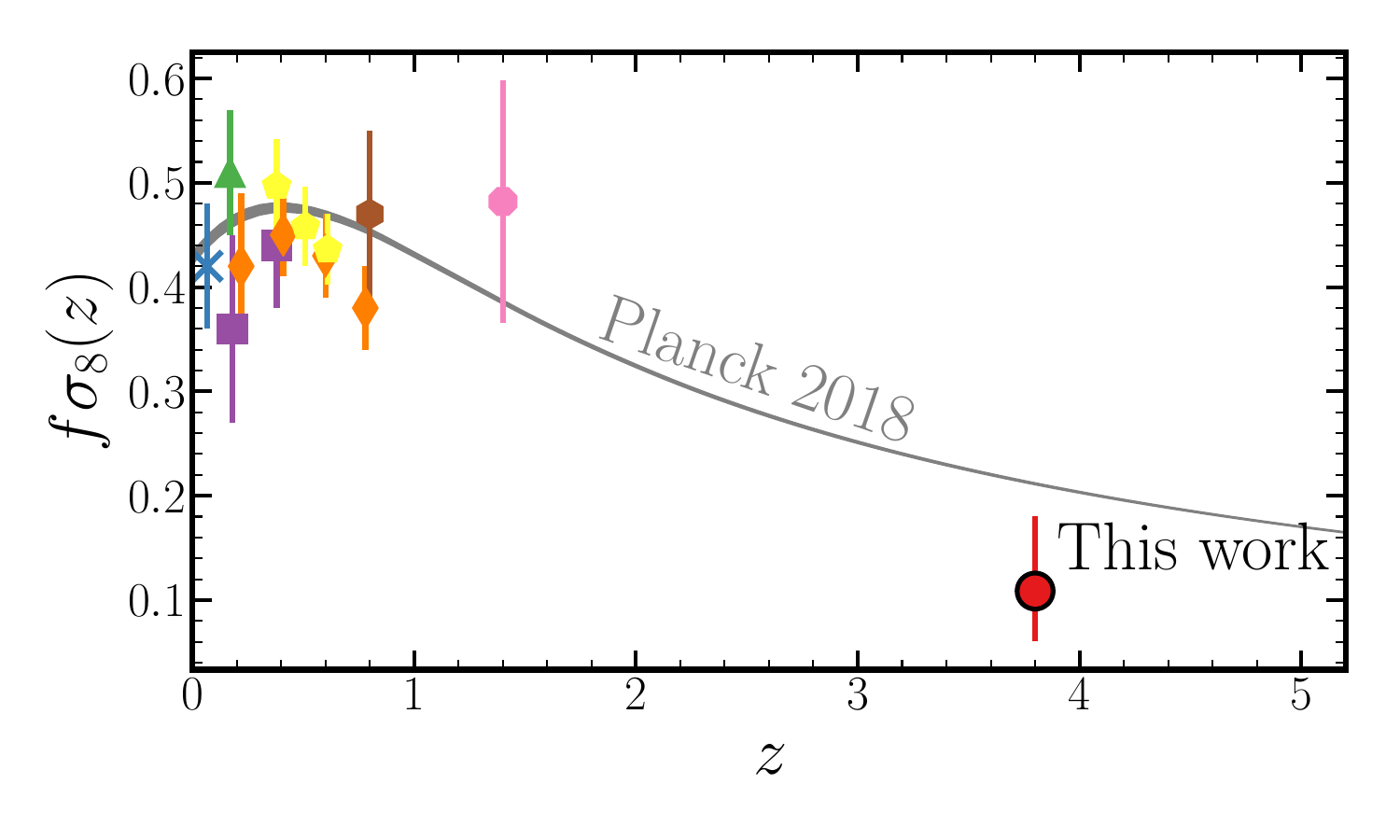}
\caption{{\it top:} Constraints on $\sigma_8$ based on 2-point statistics from our CMB lensing and galaxy-galaxy clustering measurements at $z\sim4$ (red circle) along with low-$z$ measurements from galaxy surveys. The blue cross denotes the HSC cosmic shear measurement \cite{Hamana:2020}, while the green triangle (purple square) represents the combined analysis of cosmic shear, galaxy-galaxy lensing, and projected (redshift-space) galaxy-galaxy clustering conducted with DES (KiDS, 2DFLenS, and BOSS) data \cite{DES:2018,Joudaki:2018}. The gray band shows constraints from \planck PR3 \cite{PlanckCosmology2018}. {\it bottom:} Our constraint on $f\sigma_8(z)$ (red circle) that is obtained by converting the $\Omega_{{\rm m}0}$ and $\sigma_8$ constraints into $f\sigma_8(z)$. The constraints at low-$z$ from redshift-space distortion measurements are indicated with the blue cross, green triangle, purple square, orange diamonds, yellow pentagons, brown hexagon, and magenta octagon that correspond to 6dFGRS \cite{Beutler:2012}, 2dFGRS \cite{Percival:2004}, GAMA \cite{Blake:2013}, WiggleZ \cite{Blake:2011}, BOSS \cite{Alam:2017}, VIPERS \cite{delaTorre:2013}, and FastSound \cite{Okumura:2016}, respectively.}
\label{fig:cosmological_constraints_compilation}
\end{figure}

Fig.~\ref{fig:cosmological_constraints_compilation} compares $\sigma_8$ constraints from our high-$z$ study with those from low-$z$ studies in the literature. Our constraint on $f\sigma_8(z)$ is derived by converting our constraints on $(\Omega_{{\rm m}0}, \sigma_8)$ into $f\sigma_8(z)$, where $f=-{\rm d} \ln D(z)/{\rm d} (1+z)$ is the logarithmic growth rate and $\sigma_8(z)=\sigma_8D(z)/D(0)$ is the linear matter fluctuation at redshift $z$. Our constraints on both $\sigma_8$ and $f\sigma_8(z)$ are roughly consistent with constraints computed from the Planck 2018 TT,TE,EE+lowE+lensing result \cite{PlanckCosmology2018}, while our constraints are lower than the \planck cosmology at the $1.4\sigma$ level for both $\sigma_8$ and $f\sigma_8(z)$. Motivated by the potential deviations from the \planck constraints on $\Lambda$CDM cosmology we repeat the same analysis with the assumption that the time dependent dark energy is characterized by the equation-of-state parameter $w(a)=w_0+(1-a)w_a$. We find that again our constraint is consistent with the \planck cosmology.

The HSC SSP survey will be completed by the end of 2021, which will cover $\sim$1,400 deg$^2$ of the sky and provide the GOLDRUSH sample three times as large as the one used in this paper. The Advanced Atacama Cosmology Telescope Polarimeter (AdvACT) recently released the temperature and polarization maps much deeper than {\it Planck}, and will increase sensitivity in coming years \cite{Aiola:2020,Darwish:2020}. Combining these data sets, we can significantly improve the signal-to-noise ratios of the lensing and clustering signals. In addition, the beam size of AdvACT is much smaller than {\it Planck}, which will improve systematics in CMB lensing measurements at small scales. This will enables to perform better measurements of 1-halo term. Moreover, in the near future, optical and near-infrared (O-NIR) programs of the Rubin Observatory Legacy Survey of Space and Time (LSST) \cite{LSST:2009} and the Nancy Grace Roman Space Telescope \cite{Spergel:2015} will provide much larger and cleaner LBG samples by their deeper and redder imaging. Spectroscopic observations planned with Roman will enable us to evaluate completeness and purity of the LBG samples. On the CMB side, Simons Observatory \cite{Ade:2019} and CMB-S4 \cite{Abazajian:2019}, which will provide much deeper data compared to AdvACT while keeping the beam size, are scheduled to be deployed at around the same time as the O-NIR surveys start. The combination of these data sets will allow us to extend precision cosmology studies via the structure growth from high ($z\sim 4-6$) to low redshift ($z<1$) \cite{Wilson2019}. The structure growth measurement at such a high redshift may enable us to explore a non-standard cosmological model that can resolve the $H_0$ tension such as an early dark energy model \cite{DiValentino:2021}. In fact, \citet{Klypin:2021} showed that halo clustering at $z\sim4$ ($z\sim6$) can be suppressed by $\sim20\%$ ($\sim30\%$) by early dark energy. Thus measurements of the high-redshift large scale structure can be one of the important probes for cosmology beyond the concordance model.

\vspace{3mm}
\begin{acknowledgments}
The Hyper Suprime-Cam (HSC) collaboration includes the astronomical communities of Japan and Taiwan, and Princeton University.  The HSC instrumentation and software were developed by the National Astronomical Observatory of Japan (NAOJ), the Kavli Institute for the Physics and Mathematics of the Universe (Kavli IPMU), the University of Tokyo, the High Energy Accelerator Research Organization (KEK), the Academia Sinica Institute for Astronomy and Astrophysics in Taiwan (ASIAA), and Princeton University.  Funding was contributed by the FIRST program from the Japanese Cabinet Office, the Ministry of Education, Culture, Sports, Science and Technology (MEXT), the Japan Society for the Promotion of Science (JSPS), Japan Science and Technology Agency (JST), the Toray Science  Foundation, NAOJ, Kavli IPMU, KEK, ASIAA, and Princeton University.
 
This paper makes use of software developed for the Large Synoptic Survey Telescope. We thank the LSST Project for making their code available as free software at  http://dm.lsst.org
 
This paper is based [in part] on data collected at the Subaru Telescope and retrieved from the HSC data archive system, which is operated by Subaru Telescope and Astronomy Data Center (ADC) at NAOJ. Data analysis was in part carried out with the cooperation of Center for Computational Astrophysics (CfCA), NAOJ.
 
The Pan-STARRS1 Surveys (PS1) and the PS1 public science archive have been made possible through contributions by the Institute for Astronomy, the University of Hawaii, the Pan-STARRS Project Office, the Max Planck Society and its participating institutes, the Max Planck Institute for Astronomy, Heidelberg, and the Max Planck Institute for Extraterrestrial Physics, Garching, The Johns Hopkins University, Durham University, the University of Edinburgh, the Queen’s University Belfast, the Harvard-Smithsonian Center for Astrophysics, the Las Cumbres Observatory Global Telescope Network Incorporated, the National Central University of Taiwan, the Space Telescope Science Institute, the National Aeronautics and Space Administration under grant No. NNX08AR22G issued through the Planetary Science Division of the NASA Science Mission Directorate, the National Science Foundation grant No. AST-1238877, the University of Maryland, Eotvos Lorand University (ELTE), the Los Alamos National Laboratory, and the Gordon and Betty Moore Foundation.

We thank Mathew Madhavacheril, Shuichiro Yokoyama, and Kiyotomo Ichiki for useful discussions. This work was supported in part by World Premier In-ternational Research Center Initiative (WPI Initiative), MEXT,  Japan,  and  JSPS  KAKENHI  Grant  Numbers JP15H02064, JP19H05100, JP20H01932, and JP21H00070.
\end{acknowledgments}
\bibliography{main}

\end{document}